\documentclass[a4paper,twoside,10pt]{letter}
\usepackage{graphicx,saj,multicol,subeqnarray,hyperref}

\begin{document}
\baselineskip=3.1truemm \columnsep=.5truecm
\newenvironment{lefteqnarray}{\arraycolsep=0pt\begin{eqnarray}}
{\end{eqnarray}\protect\aftergroup\ignorespaces}
\newenvironment{lefteqnarray*}{\arraycolsep=0pt\begin{eqnarray*}}
{\end{eqnarray*}\protect\aftergroup\ignorespaces}
\newenvironment{leftsubeqnarray}{\arraycolsep=0pt\begin{subeqnarray}}
{\end{subeqnarray}\protect\aftergroup\ignorespaces}

\def\degree{$^{\circ}$}
\def\kms{km~s$^{-1}$}


\markboth{\eightrm THE MAGELLANIC STREAM} 
{\eightrm S. STANIMIROVI{\' C}, J. S. GALLAGHER {\lowercase{and}} L. NIGRA}

{\ }

\publ


{\ }


\title{THE SMALL-SCALE STRUCTURE OF THE MAGELLANIC STREAM AS A FOUNDATION FOR
GALAXY EVOLUTION}


\authors{S. Stanimirovi\'c, J. S. Gallagher III and L. Nigra}

\vskip3mm


\address{Astronomy Department, University of Wisconsin, Madison, 475 North
Charter Street, \break Madison, WI 53711, USA}

\Email{sstanimi}{astro.wisc.edu}



\address{Published by the Serbian Astronomical Journal: \url{http://saj.matf.bg.ac.rs/180/pdf/001-010.pdf}}



\summary{The Magellanic Stream (MS) is the nearest example of a
gaseous trail formed by interacting galaxies. While the
substantial gas masses in these kinds of circumgalactic structures
are postulated to represent important sources of fuel for future
star formation, the mechanisms whereby this material might be
accreted back into galaxies remain unclear. Recent neutral
hydrogen (HI) observations have demonstrated that the northern
portion of the MS, which probably has been interacting with the
Milky Way's hot gaseous halo for close to 1000~Myr, has a larger
spatial extent  than previously recognized, while also containing
significant amounts of small-scale structure.  After a brief
consideration of the large-scale kinematics of the MS as traced by
the recently-discovered extension of the MS, we explore the aging
process of the MS gas through the operation of  various
hydrodynamic instabilities and interstellar turbulence.   This in
turn leads to consideration of processes whereby MS material
survives as cool gas, and yet also evidently fails to form stars.
Parallels between the MS and extragalactic tidal features are
briefly discussed with an emphasis on steps toward establishing
what the MS reveals about the critical role of local processes in
determining the evolution of these kinds of systems.
}



\begin{multicols}{2}
{


\section{1. INTRODUCTION}

Numerical simulations of galaxy formation show that galaxies can
grow by accreting gas from cosmic filaments and satellite galaxies
(e.g. Kere{\v s} et al. 2005, Dekel et al. 2009). Even at the
present time, z=0, accretion processes are likely to be ongoing
and help galaxies  to  sustain star formation (Sancisi et al.
2008, Kere{\v s} and Hernquist 2009). In any case, the response of
the gas and any associated material on approaching a giant galaxy
is an important factor in determining the fate of the accreted
matter; e.g. does infalling gas act as clouds on approximately
ballistic orbits or does it become part of a diffuse medium?
Recent studies suggest a multi-phase nature of the accreted
material: while some inflowing gas is shock-heated to near virial
temperatures, a significant amount is accreted at lower
temperatures of $T<$ few $\times$10$^5$~K where cooling times are
short.  Brooks et al. (2009) show that for a Milky Way (MW) type
galaxy, about 60-70\% of accreted gas is shocked to $T\sim10^6$ K
temperature, while both cold accretion (from cosmic filaments) and
accretion for previous mergers and satellites (`clumpy' component)
contribute each about 30-40\%. While containing a smaller amount
of gas in general, the unshocked and `clumpy' components play a
very important role for building up the disk: cold gas is
delivered close to the disk and goes on to form stars faster than
the shocked gas, which must cool before supporting star formation.

Gas stripped from satellites and by interacting galaxies are both
important sources of galactic gas infall and provide the most
direct possibilities for observing accretion processes.   As
emphasized by Kere{\v s} and Hernquist (2009), the multiphase
nature of galactic halos also enters into this issue by modifying
gas stripping processes (see also Silk et al. 1987, Gallagher and
Smith 2005, and T{\" u}llmann et al. 2006 for additional
perspectives on these issues). Understanding the astrophysics of
satellite gas accretion events thus requires knowledge of the
operation of gas stripping processes to determine the rate and
nature of gas that is injected into the surroundings as well as
models to assess the fate of the stripped gas.

We are fortunate that the Magellanic Stream (MS) offers a nearby
example of a gaseous remnant from interactions between the
Magellanic Clouds (MCs) and the Milky Way. This feature, which
extends in an arc nearly half way across the sky, offers a unique,
close-by laboratory to study physical processes of cosmological
importance in the MW halo.

In this paper, we summarize the latest observational results and
outstanding puzzles concerning the evolution of the MS gas. We
start with recent observations in Section 2. In Section 3 we
investigate the large-scale kinematics of the MS, while in Section
4 we extensively focus on physical properties of the small-scale
structure of the MS. We contrast the MS to other tidal tails in
Section 5, and conclude with a future outlook in Section 6.

\section{2. RECENT STUDIES OF THE MAGELLANIC STREAM}

The MS is a huge ($>100$ degree long) starless neutral hydrogen
(HI) structure trailing behind the MCs.  It is a remnant of the
wild past interaction of the MW with the MCs (the Large Magellanic
Cloud, LMC, and the Small Magellanic Cloud, SMC), and of the MCs
with each other. Theories have swung back and forth about the
origin of the MS, with the relative importance of tidal (Gardiner
and Noguchi 1996, Connors et al. 2006) versus ram pressure
stripping (Moore and Davis 1994, Mastropietro et al. 2005) being
still under debate. Especially challenging in recent years have
been the latest proper motion measurements (Kallivayalil et al.
2006, Piatek et al. 2008) which increased the 3-d velocity for
both Clouds, from 220 to 370 \kms for the SMC, and from 293 to 350
\kms for the LMC.
}

\end{multicols}

\centerline{\includegraphics[width=15cm]{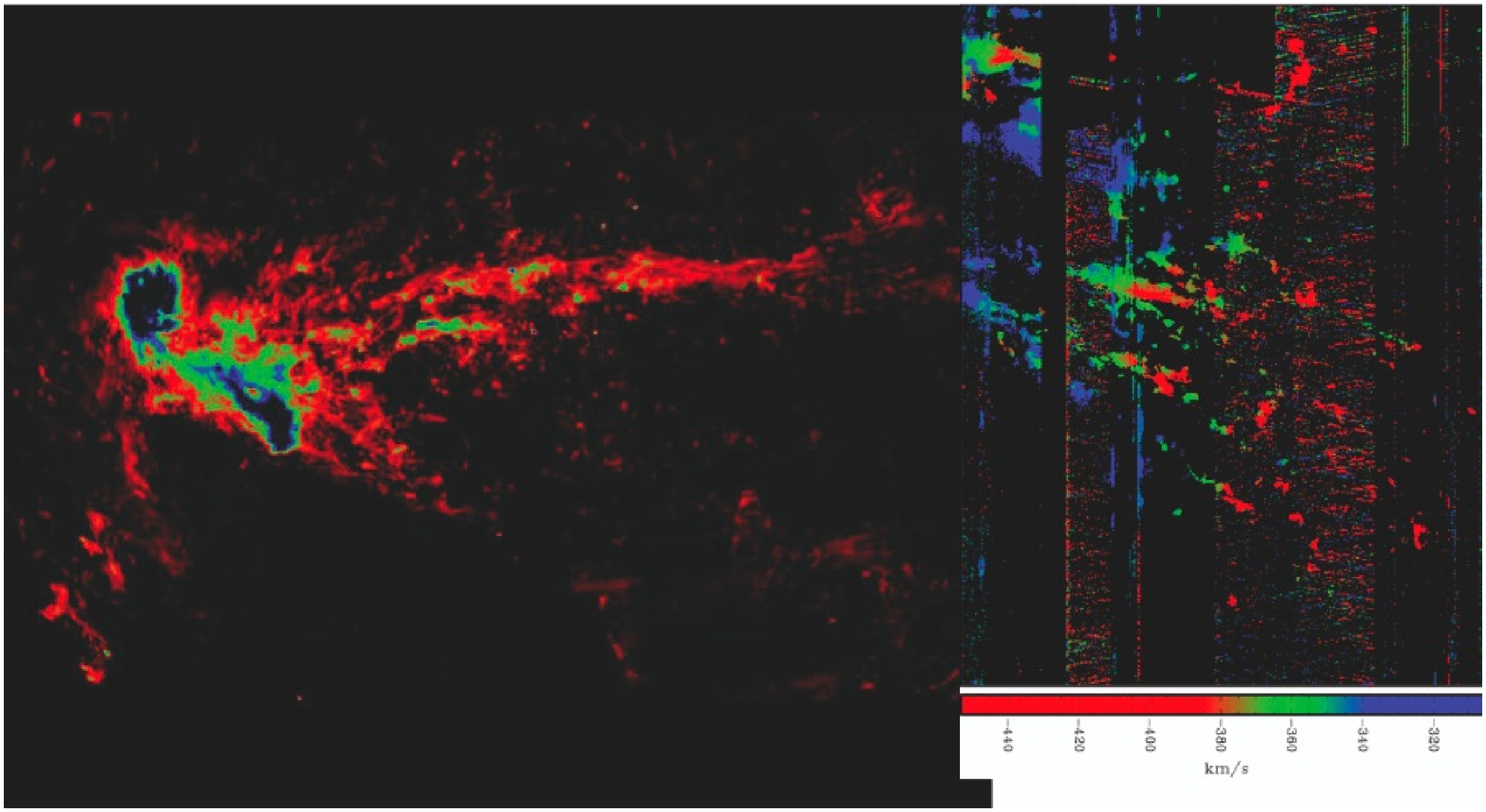}}
\figurecaption{1.}{A collage of HI observations of the MS. The
left-hand side image shows the HI column density distribution from
Putman et al. (2003) and was produced from observations obtained
with the Parkes radio telescope (angular resolution of 15$'$). The
right-hand side image shows the HI velocity field at the tip of
the MS; this image is from Stanimirovic et al. (2008) and was
obtained using Arecibo observations (angular resolution of
3.5$'$). The color bar corresponds to the Arecibo image and shows
a velocity range from $-320$ \kms (blue) to $-440$ \kms (red).}

\begin{multicols}{2}
{

After incorporating these new measurements into the orbit
calculation, Besla et al. (2007) suggested, contrary to all
previous studies, that the MCs are not likely to be bound to the
MW and may be  approaching the MW for the very first time. This
leaves essentially very little time for the SMC-LMC-MW
interactions and the formation of the MS either though tidal or
ram pressure forces. In the mean time, several observational
studies have suggested that the MW's rotational velocity is
significantly higher (254 \kms) than the IAU standard of 220 \kms
(Reid et al. 2009). This implies that the MW itself is more
massive, with a virial mass of $1.5\times10^{12}$ M$_{\odot}$.
Shattow and Loeb (2009) showed that with a more massive MW binary
orbits of the MCs are again possible and the MCs may be
gravitationally bound by the MW.

While apparently starless, the MS is the host of frequently
detected H$\alpha$ emission. The origin of this H$\alpha$ emission
has been mysterious as the expected H$\alpha$ signal excited by
the cosmic and Galactic UV background is significantly lower than
what is observed (Bland-Hawthorn and Maloney 1999). Additional
sources of ionization have been invoked, including shocks,
magnetic fields, and/or interactions between the MS clouds and the
hot halo gas.

Many past and recent low-resolution HI studies have provided
illuminating insights into the structure of the MS. A strong
velocity gradient was observed from about $+400$ \kms ~at the
location of the LMC  (Dec $\sim-68$ deg), to $-400$ \kms ~at the
tip of the MS (Dec $\sim20$\degree), the farthest away from the
MCs. Observations with the Parkes radio telescope at angular
resolution of 15$'$ (Staveley-Smith et al. 1998, Putman et al.
2003, Br{\"u}ns et al. 2005), showed interesting large-scale HI
structure in the form of two 100-degree long interwoven  filaments
(see Fig. 1, left). It was thought that the filaments become
overwhelmed by the MW halo around Dec $\sim0$\degree, ending up in
a chaotic network of small filaments and clumps. The total HI mass
of the MS is $2 \times 10^8$ M$_{\odot}$ (Putman et al. 2003). For
comparison, the HI mass of the SMC is $4.2 \times 10^8$
M$_{\odot}$ (Stanimirovic et al. 1999).

Most recent observations highlight two important observational
phenomena: the MS is significantly more extended than previously
thought, and has a significant abundance of small-scale structure.
Braun and Thilker (2004) suggested that the {\it diffuse} northern
portion of the MS extends up to Dec~40\degree, while the latest
high-resolution observations by the GALFA-HI survey (Stanimirovic
et al. 2006) showed a highly organized structure. Instead of a
chaotic HI distribution, Stanimirovi{\'c} et al. (2008) found
several filamentary structures extending up to Dec $\sim30$\degree
to the north, and reaching a heliocentric velocity of $-420$ \kms.
These filaments have a great deal of small-scale structure, mainly
in the form of discrete HI clouds, and have distinct HI
morphologies and velocity gradients (Fig. 1, right). Very
recently, Nidever et al. (2010) combined all available HI data
sets with some new Green Bank Telescope observations and showed
that there is indeed a {\it continuous} extension of the northern
MS from the areas covered by Parkes and GALFA-HI surveys all the
way to Dec~40\degree as suggested by Braun and Thilker (2004).

In summary, the present-day MS is at least 40\% longer, and about
10\% more massive, relative to the MS we knew about a few years
ago.

\vskip-2mm

\section{3. WHAT PROCESSES SHAPE THE LARGE-SCALE HI STRUCTURE OF THE MS?}

\vskip-2mm

Attempts to reproduce the observed HI morphology and velocity
gradient along the MS have had a varying level of success. For
pros and cons of various models please see Connors et al. (2006)
and Besla et al. (2008). The model-predicted HI velocity gradients
especially differ at the northern tip of the MS, making a
comparison with observations the easiest. For example, the Connors
et al. (2006) tidal model predicts a heliocentric velocity of
$\sim-400$ \kms at the extreme north tip of the MS, while the
latest gravity $+$ ram-pressure model by Mastropietro et al.
(2005) predicts $V_\mathrm{LSR} \sim-250$ \kms. In Fig. 2 we show
these two predictions with a solid and a
\vskip-3mm
\centerline{\includegraphics[height=7cm]{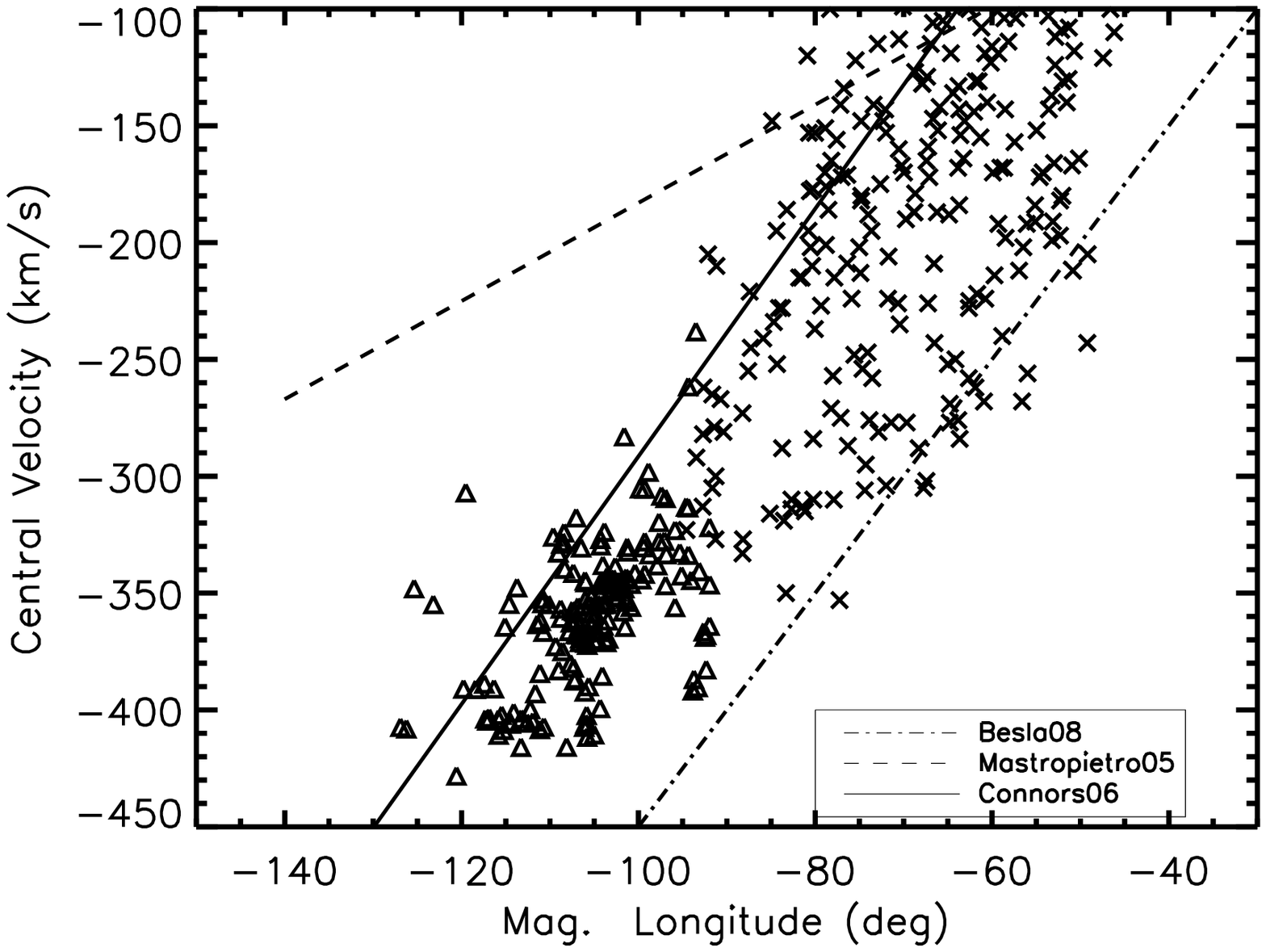}}
\vspace{-3mm}
\figurecaption{2.}{The local standard of rest (LSR) cloud velocity
as a function of Magellanic longitude. Crosses are from Putman et
al. (2003) and triangles are from Stanimirovic et al. (2008).
Three lines represent the predicted relationship from different
models: Mastropietro et al. (2005, dashed line), Connors et al.
(2006, dot-dashed line), and Besla et al. (2008, solid line).}
\noindent dashed line,
respectively. The data points represent MS clouds cataloged by
Putman et al. (2003) (crosses) and Stanimirovic et al. (2008)
(triangles).
The observed velocity of $\sim-430$ \kms by
Stanimirovic et al. (2008) is clearly in a reasonable agreement
with the purely tidal predictions. However, this is lower than the
prediction by Besla et al. (2008) of $<-500$ \kms at the extreme
MS tip (dot-dashed line in Fig. 2) obtained when the latest
proper motions of the MCs are taken into consideration in a
cosmologically-motivated gravitational model.

This comparison clearly  shows that gravity plays a major role in
the large-scale structuring of the MS kinematics. While the
updated mass of the MW (Shattow and Loeb 2009) is higher than what
has been considered in Besla et al. (2008) and may affect the MS
velocity, secondary effects like the gas drag  due to the
interaction of the MS with the halo may also be required to slow
down the MS and address details of its spatial orientation and
morphology.

\section{4. WHAT PROCESSES SHAPE THE SMALL-SCALE HI STRUCTURE OF THE MS?}

Recent Arecibo high-resolution HI observations reveal a wealth of
substructure in the MS gas, down to an angular size of $\sim3$
arcmin. Many HI clouds have multi-phase medium, while signatures
for interaction of gas structures with each other and with the
background medium are also common (Putman et al. 2003,
Bland-Hawthorn et al. 2007). The complex morphology of the HI gas,
and its observed physical properties, indicate that processes are
clearly at work on these small scales that could affect star
formation, the transfer of gas to the halo, and also may provide
additional drag affecting MS global dynamics. Although these
processes play a crucial role for gas evolution of the MS (Murray
et al. 1993, Bland-Hawthorn et al. 2007, Heitsch and Putman 2009),
it is still not clear exactly how they operate, and on what
timescales.

For example, theoretical arguments as well as simulations (Mori and
Burkert 2001, Quilis and Moore 2001, Heitsch and Putman 2009)
suggest that instabilities act on rather short timescales
($\sim100$~Myrs). This implies that the MS is being continuously
replenished, and that the stripped gas may eventually constitute a
substantial source for the MW's star formation in the form of
infalling warm ionized gas (so called  "warm drizzle",
Bland-Hawthorn et al. 2007). It is puzzling, however, that
high-resolution observations of MS HI clouds show characteristics
that indicate stability and longevity rather than rapid
destruction; they are often compact, some have a multi-phase
medium, and there are regions dense enough for the onset of
molecule formation (which we describe and discuss below).

It is important to emphasize that the global numerical simulations
rarely have resolution necessary to resolve physical processes on
small scales. For example,  simulations using smoothed particle
hydrodynamics (SPH ) can suppress development of hydrodynamic
instabilities due to smoothing (Agertz et al. 2007), while N-body
simulations ignore gas processes altogether. Grid-based modeling
can capture these small-scale instabilities better, exemplified by
simulations exploring mechanisms for excess H$\alpha$ emission in
the MS (Bland-Hawthorn et al. 2007), galaxy replenishment
(Bland-Hawthorn 2009, Heitsch and Putman 2009), and high velocity
clouds (HVCs) in the halo (Quilis and Moore 2001). It is therefore
essential to observationally constrain the effectiveness  of
various hydrodynamical instabilities for the formation and
evolution of small-scale structure.


\subsection{4.1. Cloud angular size distribution}

Stanimirovic et al. (2008) produced a catalog of HI clouds and
their basic observed properties. The cloud angular size
distribution, shown in Fig. 3, peaks at about 10$'$, while the
HI column density peaks at about $10^{19}$ cm$^{-2}$. If at a
distance of 60 kpc  (this is the distance of the SMC), then
typical clouds have a radius of $\sim100$ pc and a HI mass of
$\sim10^{3}$ M$_{\odot}$.

\vskip-3mm

\centerline{\includegraphics[width=7cm]{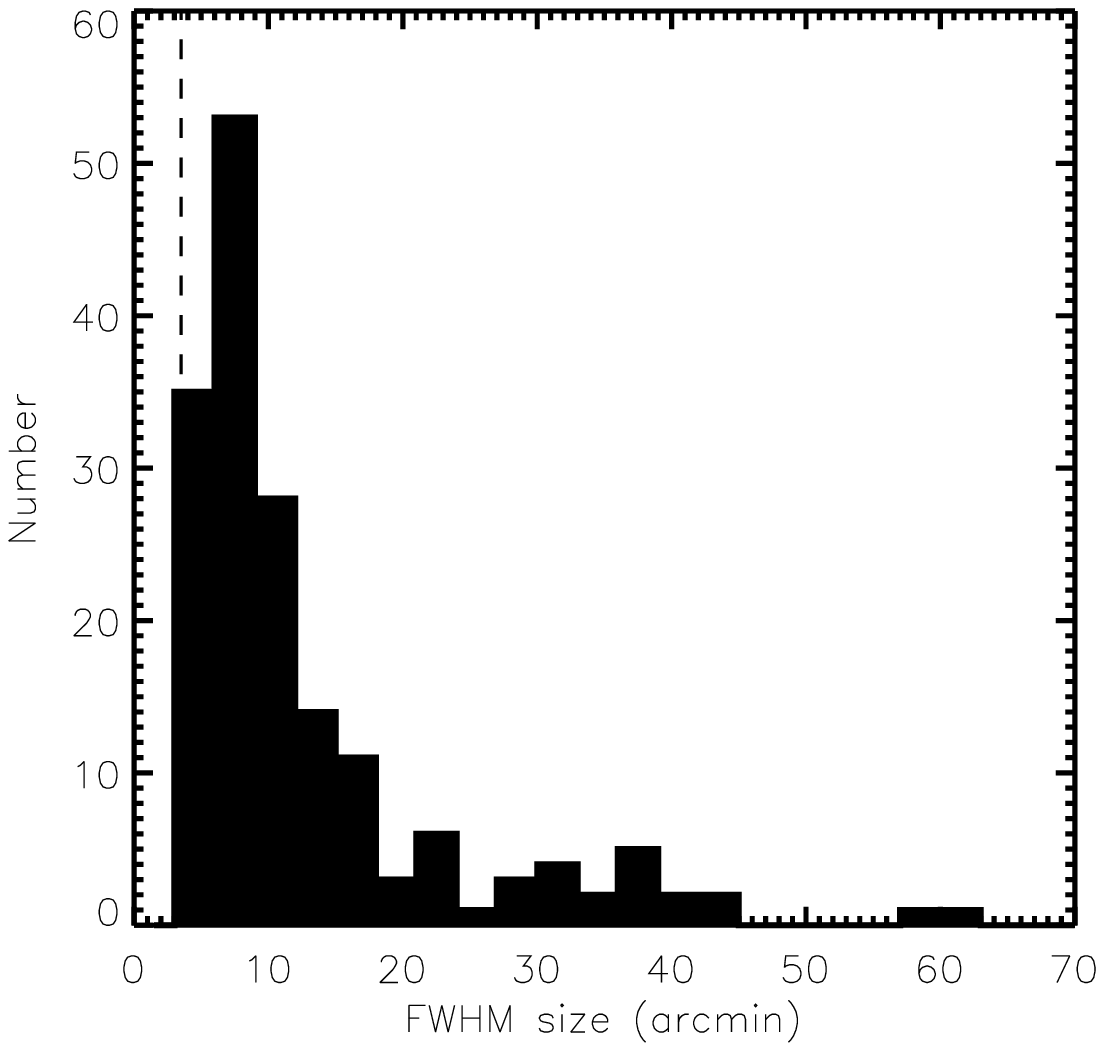}}
\vspace{-3mm}
\figurecaption{3.}{Histogram of cloud angular size (in arcmins)
for the MS cloud population from Stanimirovic et al. (2008). The
dashed line shows the angular resolution limit of the Arecibo
radio telescope.}

To investigate the importance of various hydrodynamical effects on
evolution of the MS gas we estimate their approximate timescales.

(i) Thermal instability (TI) develops due to gas cooling and would
fragment a warm stream of gas left behind  the MCs. Assuming that
the original MS had properties similar to those found in the
outskirts of the SMC, the TI fragmentation will occur on
timescales $<100$ Myrs (for details please see Stanimirovic et al.
2008).

(ii) Kelvin-Helmholtz instability (KHI) occurs at the interface
between the moving warm stream and the hot ambient medium and
provides a continuous stripping mechanism for the MS. The KHI
timescale depends on the properties of the halo gas as well
(temperature, density), which are not well constrained
observationally. However, its typical timescale is in the range of
a few hundreds to a few thousands of Myrs.

(iii) The small fragments made by TI and/or KHI are subject to the
heat transfer from the much warmer ambient medium. If undergoing
classical evaporation by heat conduction (McKee and Cowie 1977),
the HI clouds evaporate on a long timescale, $>1$ Gyr. In the case
of turbulent mixing layers the evaporation timescale would
decrease, while a magnetic field would tend to make clouds longer
lived.

To conclude,  TI and KHI must have had important effects on the
shaping of the small-scale HI structure over the MS lifetime (in
most theoretical frameworks at least 1 Gyr). While undergoing
evaporation, the HI clouds can survive for long times, and
therefore it may not be surprising to observe such clumpy
morphology at the MS tip.

If we assume that TI is the dominant shaping agent, then we can
predict a typical size of thermal fragments $\lambda_\mathrm{cool}$:
\begin{equation}
\lambda_\mathrm{cool}=\frac{k T_\mathrm{w}}{\Lambda n_\mathrm{w}}c_\mathrm{s},
\end{equation}
where, $\Lambda$ is the cooling function, $k$ is the Boltzmann
constant, $c_\mathrm{s}$ is the sound speed, and $T_\mathrm{w}$ and 
$n_\mathrm{w}$ are the
temperature and volume density of the warm neutral medium (WNM).
For the density and temperature conditions characteristic of the
SMC outskirts ($T_\mathrm{w}=8000$ K and $n_\mathrm{w}=5 \times 10^{-2}$ 
cm$^{-3}$),
"typical" thermal fragments should be about
$\lambda_\mathrm{cool}\sim200$ pc. A comparison with the peak of the
cloud angular size distribution, which corresponds to $10'$,
suggests that the MS tip is at a distance of $\sim70$ kpc. While
this simple, back-of-the-envelope calculation is only
demonstrative, it is interesting that our distance estimate agrees
well with the predictions from tidal models (e.g. Connors et al.
2006). Even more impressively, our distance estimate is in
agreement with the recent estimate of 75 kpc based on a model by
Jin and Lynden-Bell (2008). This model assumed that energy and
angular momentum are conserved along the MS, and that the MS is
trailing on a planar orbit around the Galactic center.

\subsection{4.2. Multi-phase medium}

Another interesting phenomenon is the multi-phase HI structure of
the MS. About 15\% of clouds in the sample of Stanimirovic et al.
(2008) have velocity profiles whose fits require two temperature
components. This suggests the existence of a multi-phase medium at
a significant distance from the MW plane. We find evidence for
warm gas, with a velocity FWHM of about 25 \kms, and a cooler
component, with a FWHM generally in the range 3-15 \kms.
Similarly, Karberla and Haud (2006) investigated velocity profiles
along the MS based on the Leiden/Argentine/Bonn data (Kalberla et
al. 2005). They found that 27\% of MS profiles at positive LSR
velocities (close to MCs), and 12\% of profiles at negative LSR
velocities, require two temperature components.

In addition, Matthews et al. (2009) detected the first HI
absorption lines against radio background sources in the direction
of the MS close to the MCs.  The two detected absorption features
have a velocity FWHM of 4.2 and 5.0 \kms, respectively. The spin
temperature of the absorbing clouds of 80 and 70 K was derived,
resulting in the HI column density of $\sim2 \times 10^{20}$
cm$^{-2}$. The only direct detection of H$_2$ in absorption is by
Richter et al. (2001),  who used Far Ultraviolet Spectroscopic
Explorer (FUSE), and found an excitation temperature of $\sim140$
K and $N(\mathrm{H}_2) = 3 \times 10^{16}$ cm$^{-2}$.

Clearly, the MS contains a multi-phase medium. This is exciting as
some cold cores are reaching temperature and HI column densities
usually required for molecule (CO) formation, providing potential
for future star formation. One question that remains is whether
these cold cores were formed in the MS, or have been stripped from
the MCs.

\centerline{\includegraphics[width=7.2cm]{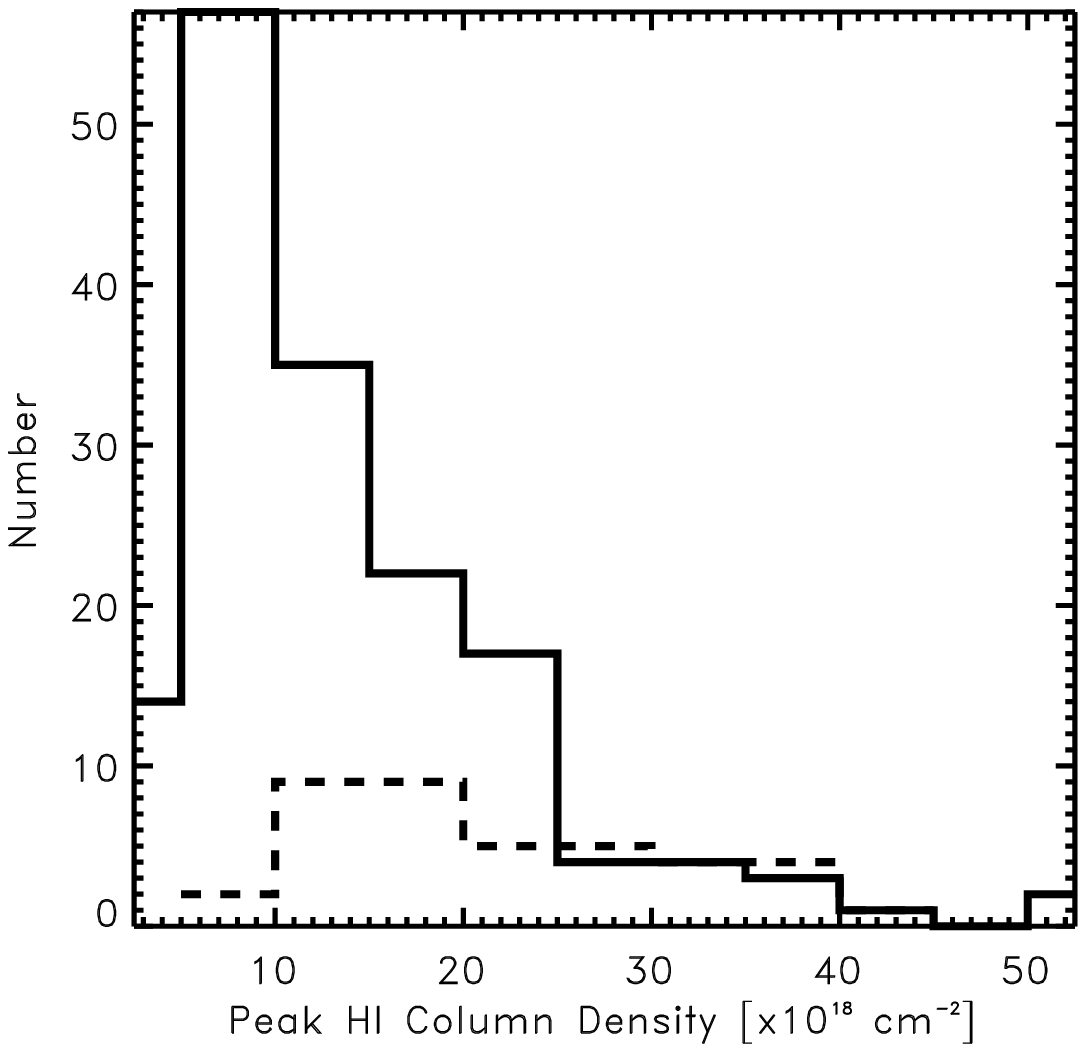}}
\vspace{-3mm}
\figurecaption{4.}{Histograms of the peak HI column density (in
$10^{18}$ cm$^{-2}$) for HI clouds with a single velocity
component (solid line) and HI clouds with the multi-phase
structure (dashed line). The 3-$\sigma$ sensitivity limit of the
GALFA-HI survey used for this study is $\sim3 \times 10^{18}$
cm$^{-2}$.}
In Fig. 4 we show the HI column density of single-phase and
multi-phase clouds at the tip of the MS.
While single-phase
clouds peak at $N(\mathrm{HI}) \sim 10^{19}$ cm$^{-2}$, the multi-phase
clouds have  higher column densities $N(\mathrm{HI}) \sim 1.5-4 \times
10^{19}$ cm$^{-2}$. The two distinctly different hystograms
suggest intrinsic difference between single- and multi-phase
clouds, with the latter one being found in better shielded, more
condensed regions.

The level of turbulent motions of colder cores with respect to the
warmer envelopes can be gauged by calculating the sonic Mach
number of cold cores: $M=|V_\mathrm{c} -
V_\mathrm{w}|/\mathrm{FWHM}_\mathrm{w}$, with $V_\mathrm{c}$ and 
$V_\mathrm{w}$
being velocity centroids of the cold and warm cloud components,
and $\mathrm{FWHM}_\mathrm{w}$ being the velocity FWHM of the warm component. The
histogram of $M$ values is shown in Fig. 5; $\sim90$\% of data
points are within $|M|<1.5$ suggesting subsonic or mildly
supersonic motions.   For comparison, Heiles and Troland (2003)
found supersonic internal motions for the MW cold neutral medium
(CNM) clouds with $M\sim3$, while Kalberla and Haud (2006) found
that most HVCs have $M\sim1.5$ for cold cores relative to their
warm envelopes.
%
%
%
%

\vspace{-.6cm}

\centerline{\includegraphics[width=7cm]{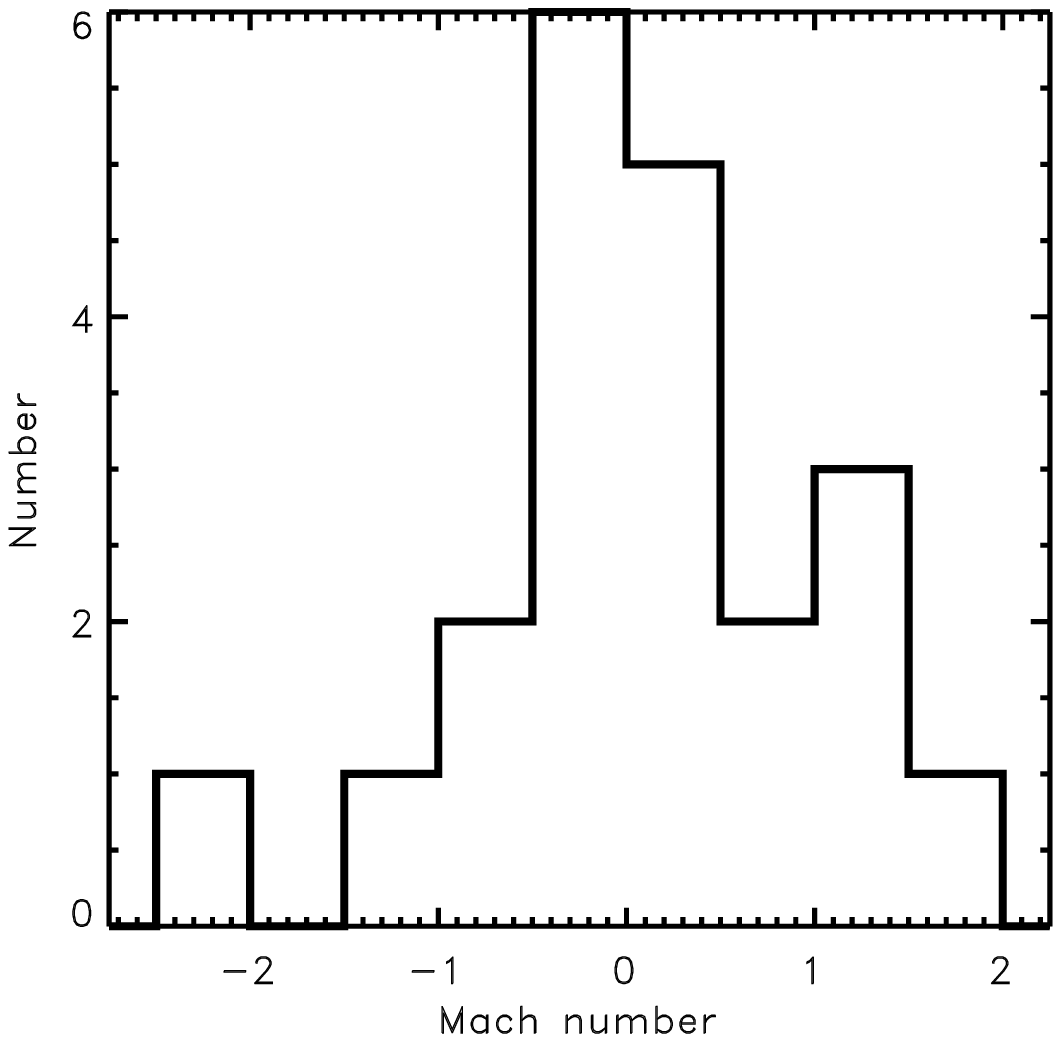}}
\vspace{-3mm}
\figurecaption{5.}{Histogram of the sonic Mach number of  MS cold
HI cloud cores relative to warm cloud envelopes. }

The existence of a multi-phase medium in the MW halo  at a
distance of $\sim60$ kpc is however not expected from a
theoretical point of view. Based on the consideration of
cooling/heating processes, Wolfire et al. (1995) suggested that
pressure-confined multi-phase clouds should not be found in the MW
halo at distances larger than 25 kpc. The thermal pressure
required for the co-existence of the CNM and WNM at a distance of
45 kpc from the MW plane is $>100$ K cm$^{-3}$, while a
plane-stratified isothermal MW halo with $T=10^6$ K provides a
thermal pressure of $<100$ K cm$^{-3}$. While Sternberg et al.
(2002) showed that dark matter can provide an additional
confinement mechanism, thereby allowing the existence of
multi-phase structure at distances $<150$ kpc, the amount of dark
matter associated with the MS is unknown. Re-consideration of halo
properties and/or requirements for the existence of the
multi-phase medium are clearly needed. For example, it is becoming
well-accepted that dynamical events such as collisions between
turbulent flows can initiate fast cycling of interstellar gas
between multiple phases.

While our discussion above focused only on the  HI multi-phase
structure, further insights into the rich multi-phase structure in
the MW halo have been provided by UV and optical absorption
studies probing various ionization states. Detections of OVI
absorption from gas associated with the MS by Sembach et al.
(2003) give strong support for the existence of an ionized
component around the MS with $T<10^6$ K. It is generally
interpreted that this component represents an interface between
the hot halo gas ($T\sim 10^6$ K) and the cooler MS gas.

Studies of lower ionization states suggest the existence of
diffuse envelopes of somewhat cooler, partially ionized gas that
is not visible in current HI surveys. Specifically, SiIII provides
a more sensitive probe of  ionized components of the infalling
gas,  likely probing different phases than OVI with $T=10^{4-4.5}$
K and the corresponding $N_\mathrm{HI}$ $\sim 10^{17}$~cm$^{-2}$, well
below current HI survey sensitivity (Shull et al. 2009). Such gas
has been detected along a sightline in the northern MS with
velocities associated with the MS (Collins et al. 2009). Fox et
al. (2010, in prep) have performed an analysis of many low- and
high-ion species in UV and optical absorption, including SiIII,
against background sources in the northern MS.  From this work, a
picture emerges of a diffuse, multi-phase transition structure
between the warm, mostly neutral envelope gas detected in HI and
the hot, mostly ionized envelope gas detected in OVI.

\subsection{4.3. Turbulence as a generator of small-scale\hfill\ \break
$\hphantom{\bf 4.3.\ }$ structure?}

Instead of being predominantly formed out of the smooth diffuse
WNM, the clumpy and multi-phase small-scale structure in the MS
could be a result of turbulent inhomogeneities that originated in
the MCs and were simply stripped during the MW-MCs interactions.
Also, an alternative to hydrodynamical instabilities being the
main driving source (as considered in previous sections), are
large-scale shearing and tidal flows that can induce turbulence on
large scales and provide an energy cascade and formation of
structure on smaller scales. We briefly explore these
possibilities in this section.

It has been shown that the HI distribution in the SMC and the LMC
is turbulent and can be described with a spatial power spectrum
$P(k) \propto k^{-\gamma}$ (Stanimirovic et al. 1999, Elmegreen et
al. 2001). The power-law slope of the density field is
$\gamma=3.4$ in the case of the SMC (Stanimirovic and Lazarian
2001), and $\gamma=3.7$ for the LMC (Elmegreen et al. 2001). One
puzzling issue, however, is that these power spectra do not show
significant changes at the largest sampled scales. This could be
interpreted as evidence for interstellar turbulence being driven
on scales larger than the size of the SMC/LMC, $> 4-5$ kpc.

Recently, Burkhart et al. (2010) developed a new method to gauge
the spatial variations of turbulence by applying the 3rd and 4th
statistical moments (or skewness and kurtosis) on the observed HI
column density distribution. Based on MHD simulations, these
high-order statistical moments are well correlated with the sonic
Mach number. Therefore, by measuring skewness and kurtosis for the
observed data, we can use the correlations derived from simulated
data sets to retrieve the spatial distribution of the sonic Mach
number, which provides an estimate for the local level of
turbulence. Burkhart et al. (2010) applied this method to the HI
column density image of the SMC and found that regions with the
highest level of turbulence are located at the boundaries of the
SMC bar. This suggests that large-scale motions between the bar
and the surrounding diffuse HI, possibly induced by tidal
interactions between the SMC, the LMC and the MW, or some kind of
shearing flows, may be imprinting a strong energy signature on the
HI gas.

Similarly, Goldman (2000) suggested that the HI turbulence in the
SMC was induced by large-scale flows from tidal interactions with
the MW and the LMC about $2\times 10^8$ yrs ago. Such large-scale
bulk flows could have generated turbulence through shear
instabilities. If shearing flows were able to leave such strong
imprint on the SMC, they must be also affecting the MS gas as well
and may be responsible for the formation of the small-scale
structure we observe in HI.

The need for an initially clumpy MS gas has also been highlighted
recently by Bland-Hawthorn et al. (2007) who proposed a
shock-cascade  process to explain the observed H$\alpha$ emission
along the MS.  Two most important aspects of this study are: an
initially clumpy distribution of the MS gas, and a strong
interaction between the MS clouds and the MW halo which drives the
collisionally excited H$\alpha$ emission. As the MS clouds
upstream experience gas ablation by the oncoming hot Galactic
wind, the ablated gas is slowed down and transported behind the
clouds. The ablated gas further collides with the clouds
downstream, resulting in shock ionization of HI clouds. This
shock-cascade model can explain measured H$\alpha$ intensities
along the MS.
The shock-cascade model predicts that large changes should take
place in the HI distribution on timescales of 100-200 Myrs. The
ablation process erodes the low density HI gas, slowly eating into
the high-density regions. As a result, after about 200 Myrs, the
HI column density distribution is highly asymmetric (see Fig.
6): it peaks at $N(\mathrm{HI})\sim10^{19}$ cm$^{-2}$ and is missing both
low-  and high-density gas relative to the initial HI
distribution. As the tip of the MS has been exposed to the halo
the longest, the shock-cascade process should be clearly
noticeable here.

In Fig. 6 we compare the observed HI column density probability
density function (PDF) with the  same quantity at two snap-shots
in the Bland-Hawthorn et al.'s simulation: 70  and 270 Myrs after
the initial exposure of the MS to the halo wind (shown as dashed
and dot-dashed lines in Fig. 6). The observed PDF was derived by
taking the data from Stanimirovic et al. (2008), deriving the HI
column density distribution, and simply dividing this by 5 to
account for the difference in the areas  probed by observations
and the simulation (the simulated area is about 5 times smaller
than that probed by observations). The large difference in the
simulated data after 200 Myrs  is clearly visible, and the later
distribution is missing both low- and high-density gas.

\vskip1mm

\centerline{\includegraphics[height=6.5cm]{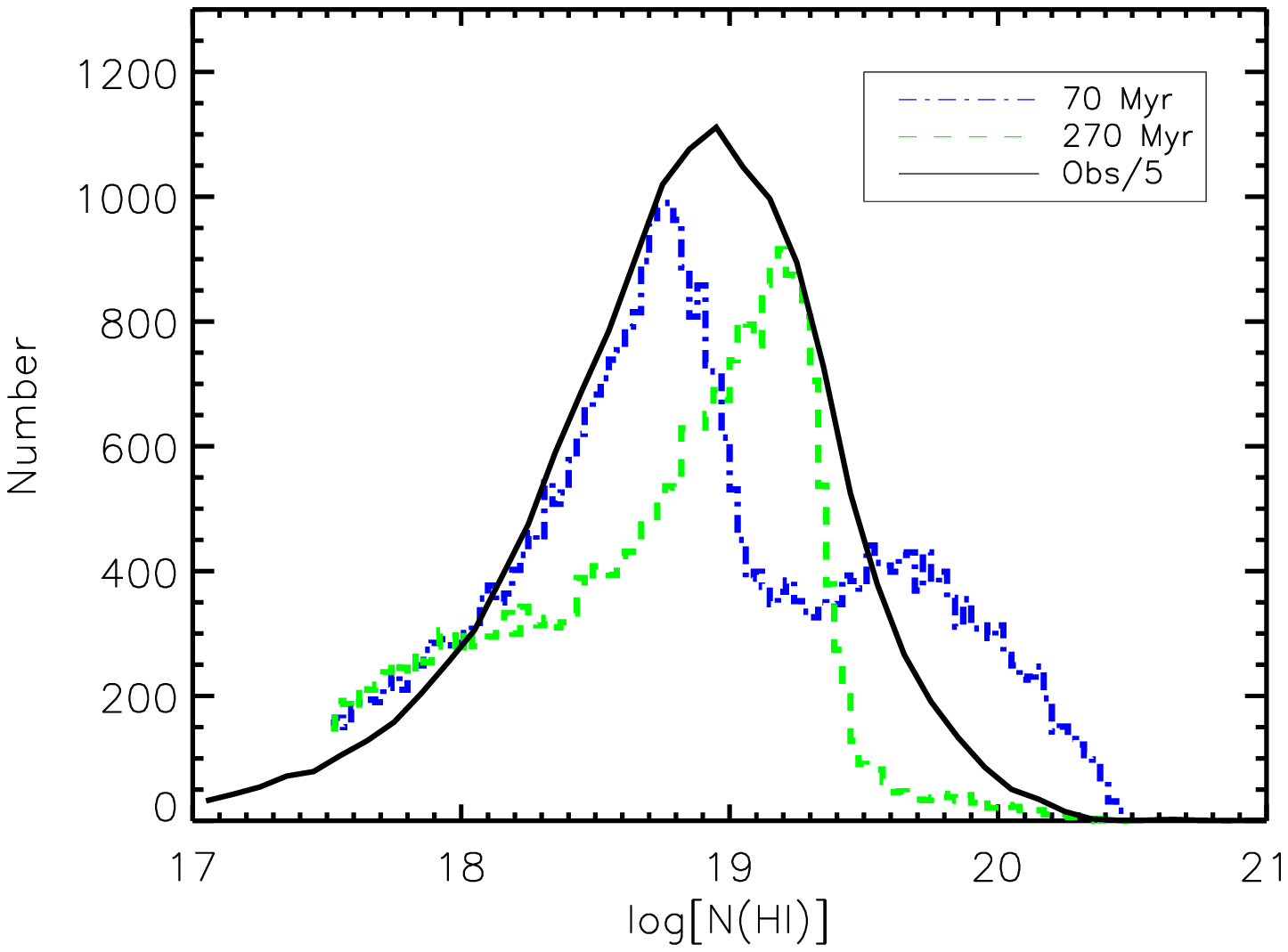}}
\vspace{-1mm}
\figurecaption{6.}{Evolution of the HI column density PDF in the
shock-cascade simulation by Bland-Hawthorn et al. (2007). The blue
dot-dashed and green dashed lines shows times stamps in the
simulation at 70 and 270 Myrs, respectively. The solid line shows
the HI column density PDF derived by using observations from
Stanimirovic et al. (2008). The observed PDF was divided by 5
because observations sample five times larger projected area than
the simulation.}

However, the observed PDF is not similar to any of the simulated
PDFs. Contrary to a highly asymmetric simulated $N(\mathrm{HI})$ PDF, the
observed PDF is highly symmetric and almost Gaussian. It clearly
contains more low- and high-density gas than the end point of the
simulation. As shown in Burkhart et al. (2010), subsonic
turbulence produces Gaussian column density PDFs, while in the
case of supersonic turbulence PDFs are highly skewed. This again
highlights the difference between observations and the simulation:
simulated distributions appear significantly more turbulent than
what observations show. This suggests that the proposed ablation
process is too fast and something must be slowing it down and
helping the MS clouds survive longer.

The structure of the boundary between clouds and the hot
atmosphere of the MW is one factor which could account for the
apparently slow rate of mass ablation in the MS.  Numerical models
by Vieser and Hensler (2007) indicate that while conductive heating
promotes the evaporation of a cooler, dense cloud moving with
respect to a rarified hot medium, it also can significantly reduce
the action of the KTI.  Since under these circumstances KTI is
likely to be the dominant mode of cool cloud disruption, the net
effect of conductive heating then is to substantially extend cloud
lifetimes. This type of mechanism also is consistent with the
presence of gas with a range of ionization potentials in the MS,
and merits further exploration.

\section{5. THE MS AS A VEHICLE FOR UNDERSTANDING CONDITIONS FOR STAR FORMATION
IN TIDAL TAILS IN GENERAL}

The evolution of gas-rich tidal debris involves processes
operating on a variety of density, temperature and mass scales.
Among these star formation is one interesting end point for small
scale structures. While massive tidal tails, such as those in
some of the Arp interacting galaxies, show clear evidence for active
star formation (e.g. Schombert et al. 1990, Gallagher et al.
2001), other gas-rich tidal features, such as the eastern HI arm
of M51 (Rots et al. 1990), remain as purely gaseous structures.
However, the increased sensitivity of \ recent observations \ are re-
%
%
%
%
\noindent vealing star formation under a wider range of conditions than
previously recognized.

On larger scales, these include the production of relatively large
concentrations of gas that can become tidal dwarf galaxies and are
capable of supporting extensive star formation (e.g. Mirabel et
al. 1992, Bournaud and Duc 2006, Weilbacher et al. 2000, Smith et
al. 2010 and references therein). A key point beyond this was the
recognition that even relatively diffuse extragalactic tidal gas
systems can contain significant amounts of molecular gas (Walter
and Heithausen 1999, Braine et al. 2001, Taylor et al. 2001).
Perhaps it then is to be expected that star formation on small
scales in tidal debris also is showing up, especially in deep
images obtained in the optical or in the ultraviolet with Galaxy
Evolution Explorer (GALEX) (e.g. de Mello et al. 2008, Werk et al.
2008, Thilker et al. 2009).

The MS, unlike the Magellanic bridge (Mizuno et al. 2006,
Nishiyama et al. 2007), is not known to contain classical
molecular clouds or candidate star forming regions. Thus the MS
appears to be a large scale example of an HI stream that is
sterile against star formation. How then does the MS differ from
extragalactic HI features, including many tidal tails,  that
support star formation? Can studies of the MS provide insights
into why the observed gas clumps evidently do not grow and become
gravitationally unstable and yet also survive for long times? The
low mean column density is likely to be one factor and dust
content may be another. As discussed by Maybhate et al.  (2007)
and Boquien et al. (2009), star formation usually is observed when
$N(H)>3 \times 10^{20}$~cm$^{-2}$, which is larger than the
$N(HI)$ seen in most of the stream.

However, since the gas in the MS can be studied in considerable
detail, it should be possible to go beyond this type of important
but empirical bulk diagnosis, and carry out detailed analysis of
the evolution of typical MS clumps.  For example, as we discussed
earlier, the Bland-Hawthorn et al. (2007) provides useful initial
predictions for the evolution of the column density distribution
within the MS. Comparisons between even more sophisticated models
and the new MS observations can be expected to yield insights not
only into the astrophysics controlling the MS, but more generally
into the fate of HI injected into the vicinities of galaxies by
interactions or other cosmologically related processes (e.g.
Kere{\v s} and Hernquist 2009).

\vskip-2mm

\section{6. CONCLUSIONS, ON-GOING AND FUTURE STUDIES}

\vskip-2mm

With the improved resolution and sensitivity of radio telescopes,
abundant and rich small-scale HI structure in the MS is being
revealed even in regions located the farthest away from the parent
MCs and deeply embedded in the hot MW halo. The HI clouds often
show multi-phase signatures and appear shielded from high
turbulence caused by various types of gas stripping by the MW
halo. Occasionally, even cold cores with column densities that
could support molecule formation are found.

What physical processes produce such rich structure in a tidal
tail like the MS, how will this structure evolve as it interacts
with the surrounding hot medium, and will it eventually infall to
the MW disk? These questions have led us to explore the importance
of various hydrodynamical instabilities and their effectiveness.
Our analytical consideration of timescales, as well as recent
numerical advances, suggest that thermal and Kelvin-Helmholtz
instabilities operate on timescale much shorter than the MS
formation time and hence must be important. Indirect studies of
turbulence in the MCs suggest that large-scale shearing/tidal
flows may be able to drive turbulence and cascade to smaller
scales. Yet, the observed HI column density distribution and
highly rich temperature structure observed in the MS over a range
from $\sim100$ to $\sim10^5$ K, paint a picture of stable,
long-lived  environments. One promissing solution could be the
nature of boundary regions between the MS clouds and the MW halo.

To investigate the role of the MW halo in shaping of the
small-scale MS structure, we are in the process of placing
constraints, in the radio and optical regimes, on characteristics
of the gas in the transition region from the MS gas to the ambient
MW halo medium. Using the National Radio Astronomy Observatory
Green Bank Telescope we have obtained the most sensitive HI
emission images of portions of the MS to date and have begun to
analyze the structure and kinematics of the gas. Preliminary
results from this study can be found in Nigra et al. (2009).

In the near future, a large step forward in understanding the MS
properties will come from the Galactic Australian Square Kilometer
Array Pathfinder GASKAP project. GASKAP is a study of the 21-cm
line of HI and the 18-cm lines of OH in the Galactic Plane and the
Magellanic Clouds and Stream using a new radio interferometer
(ASKAP, Johnston et al. 2007) under development in Australia.
ASKAP will consist of 36 12-m antennas, each with a focal plane
phased array, and operating over a frequency range 700 MHz to 1.8
GHz. This new instrument is expected to become operational in late
2012. GASKAP images will have an order of magnitude higher angular
resolution ($\sim20"$) and sensitivity ($\sim0.04$ K) relative to
any previous large-scale survey of the MS and will constrain
various eroding processes shaping the MS.


\acknowledgements{It is a pleassure to thank Joss Bland-Hawthorn
for stimulating discussions and sharing of simulations. We
acknowledge support by the NSF grant AST-0908134 to SS and
AST-0708967 to JSG. SS also thanks the Research Corporation for
Science Advancement for their support.}
}

\end{multicols}

\vskip 5mm
\references

\begin{multicols}{2}
{

\vskip-3mm

Agertz, O. et al.: 2007, \journal{Mon. Not. R. Astron. Soc.},
\vol{380}, 963.

Besla, G., Kallivayalil, N., Hernquist, L., Robertson, B., Cox,
T.~J., van der Marel, R.~P. and Alcock, C.: 2007,
\journal{Astrophys. J.}, \vol{668}, 949.

Besla, G., Kallivayalil, N., Hernquist,, L., vander Marel, R.~P.,
Cox, T.~J., Robertson, B. and Alcock, C.: 2008, \journal{IAU
Symp.}, \vol{256}, 119.

Bland-Hawthorn, J. and Maloney, P.~R.: 1999, \journal{Astron. Soc. Pacific. 
Conference Series}, \vol{166}, 212.

Bland-Hawthorn, J., Sutherland, R., Agertz, O. and Moore, B.:
2007, \journal{Astrophys. J.}, \vol{670}, L109.

Bland-Hawthorn, J.: 2009, \journal{IAU Symp.}, \vol{254}, 241.

Boquien, M. et al.: 2009, \journal{Astron. J.}, \vol{137}, 4561.

Bournaud, F. and Duc, P.-A.: 2006, \journal{Astron. Astrophys.},
\vol{456}, 481.

Braine, J., Duc, P.-A., Lisenfeld, U., Charmandaris, V., Vallejo,
O., Leon, S. and Brinks, E.: 2001, \journal{Astron. Astrophys.},
\vol{378}, 51.

Braun, R. and Thilker, D.~A.: 2004, \journal{Astron. Astrophys.},
\vol{417}, 421.

Brooks, A.~M., Governato, F., Quinn, T., Brook, C.~B. and Wadsley,
J.: 2009, \journal{Astrophys. J.}, \vol{694}, 396.

Br{\"u}ns, C. et al.: 2005, \journal{Astron. Astrophys.},
\vol{432}, 45.

Burkhart, B., Stanimirovi{\'c}, S., Lazarian, A. and Kowal, G.:
2010, \journal{Astrophys. J.}, \vol{708}, 1204.

Collins, J.~A., Shull, J.~M. and Giroux, M.~L.: 2009,
\journal{Astrophys. J.}, \vol{705}, 962.

Connors, T.~W., Kawata, D. and Gibson, B.~K.: 2006, \journal{Mon.
Not. R. Astron. Soc.}, \vol{371}, 108.

de Mello, D.~F., Smith, L.~J., Sabbi, E., Gallagher, J.~S.,
Mountain, M. and Harbeck, D.~R.: 2008, \journal{Astron. J.},
\vol{135}, 548.

Dekel, A. et al.: 2009, \journal{Nature}, \vol{457}, 451.

Elmegreen, B.~G., Kim, S. and Staveley-Smith, L.: 2001,
\journal{Astrophys. J.}, \vol{548}, 749.

Fox, A.~J. et al.: 2010, in preparation.

Gallagher, S.~C., Charlton, J.~C., Hunsberger, S.~D., Zaritsky,
D. and Whitmore, B.~C.: 2001, \journal{Astron. J.}, \vol{122}, 163.

Gallagher, J.~S. and Smith, L.~J.: 2005, \journal{Astron. Soc. Pacific. 
Conference Series}, \vol{331}, 147.

Gardiner, L.~T. and Noguchi, M.: 1996, \journal{Mon. Not. R.
Astron. Soc.}, \vol{278}, 191.

Goldman, I.: 2000, \journal{Astrophys. J.}, \vol{541}, 701.

Heiles, C. and Troland, T.~H.: 2003, \journal{Astrophys. J. Suppl.
Series}, \vol{145}, 329.

Heitsch, F. and Putman, M.~E.: 2009, \journal{Astrophys. J.},
\vol{698}, 1485.

Jin, S. and Lynden-Bell, D.: 2008, \journal{Mon. Not. R. Astron.
Soc.}, \vol{383}, 1686.

Johnston, S. et al.: 2007, \journal{Publ. Astron. Soc. 
Australia}, \vol{24}, 174.

Kalberla, P.~M.~W., Burton, W.~B., Hartmann, D., Arnal, E.~M.,
Bajaja, E., Morras, R., P{\"o}ppel, W.~G.~L.: 2005,
\journal{Astron. Astrophys.}, \vol{440}, 775.

Kalberla, P.~M.~W. and Haud, U.: 2006, \journal{Astron.
Astrophys.}, \vol{455}, 481.

Kallivayalil, N., van der Marel, R.~P., Alcock, C., Axelrod, T.,
Cook, K.~H., Drake, A.~J. and Geha, M.: 2006, \journal{Astrophys.
J.}, \vol{638}, 772.

Kere{\v s}, D., Katz, N., Weinberg, D.~H. and Dav{\'e}, R.: 2005,
\journal{Mon. Not. R. Astron. Soc.}, \vol{363}, 2.

Kere{\v s}, D. and Hernquist, L.: 2009, \journal{Astrophys. J.}, \vol{700}, L1.

Mastropietro, C., Moore, B., Mayer, L., Wadsley, J. and Stadel,
J.: 2005, \journal{Mon. Not. R. Astron. Soc.}, \vol{363}, 509.

Matthews, D., Staveley-Smith, L., Dyson, P. and Muller, E.: 2009,
\journal{Astrophys. J.}, \vol{691}, L115.

Maybhate, A., Masiero, J., Hibbard, J.~E., Charlton, J.~C., Palma,
C., Knierman, K.~A. and English, J.: 2007, \journal{Mon. Not. R.
Astron. Soc.}, \vol{381}, 59.

McKee, C.~F. and Cowie, L.~L.: 1977, \journal{Astrophys. J.},
\vol{215}, 213.

Mirabel, I.~F., Dottori, H. and Lutz, D.: 1992, \journal{Astron.
Astrophys.}, \vol{256}, L19.

Mizuno, N., Muller, E., Maeda, H., Kawamura, A., Minamidani, T., Onishi, T., 
Mizuno, A. and Fukui, Y.: 2006, \journal{Astrophys. J.}, \vol{643}, L107. 

Moore, B. and Davis, M.: 1994, \journal{Mon. Not. R. Astron.
Soc.}, \vol{270}, 209.

Mori, M. and Burkert, A.: 2001, \journal{Astron. Soc. Pacific. Conference 
Series}, 222, 359.

Murray, S.~D., White, S.~D.~M., Blondin, J.~M. and Lin, D.~N.~C.:
1993, \journal{Astrophys. J.}, \vol{407}, 588.

Nidever et al.: 2010, in preparation.

Nigra, L., Stanimirovic, S., Gallagher, J.~S., III, Lockman,
F.~J., Nidever, D.~L. and Majewski, S.~R.: 2009,
\journal{arXiv.org}, 0908.4218.

Nishiyama, S. et al.: 2007, \journal{Astrophys. J.}, \vol{658},
358.

Piatek, S., Pryor, C. and Olszewski, E.~W.: 2008, \journal{Astron.
J.}, \vol{135}, 1024.

Putman, M.~E., Staveley-Smith, L., Freeman, K.~C., Gibson, B.~K.
and Barnes, D.~G.: 2003, \journal{Astrophys. J.}, \vol{586},
170.

Quilis, V. and Moore, B.: 2001, \journal{Astrophys. J.},
\vol{555}, L95.

Reid, M.~J. et al.: 2009, \journal{Astrophys. J.}, \vol{700}, 137.

Richter, P., Sembach, K. R., Wakker, B. P. and Savage, B. D.:
2001, \journal{Astrophys. J.}, \vol{562}, L181.

Rots, A.~H., Bosma, A., van der Hulst, J.~M., Athanassoula, E. and
Crane, P.~C.: 1990, \journal{Astron. J.}, \vol{100}, 387.

Sancisi, R., Fraternali, F., Oosterloo, T. and van der Hulst, T.:
2008, \journal{Astron. Astrophys. Reviews}, \vol{15}, 189.

Schombert, J.~M., Wallin, J.~F. and Struck-Marcell, C.: 1990,
\journal{Astron. J.}, \vol{99}, 497.

Sembach, K.~R. et al.: 2003, \journal{Astrophys. J. Suppl. Series},
\vol{146}, 165.

Shattow, G. and Loeb, A.: 2009, \journal{Mon. Not. R. Astron.
Soc.}, \vol{392}, L21.

Shull, J.~M., Jones, J.~R., Danforth, C.~W. and Collins, J.~A.:
2009, \journal{Astrophys. J.}, \vol{699}, 754.

Silk, J., Wyse, R.~F.~G. and Shields, G.~A.: 1987,
\journal{Astrophys. J.}, \vol{322}, L59.

Smith, B.~J., Giroux, M.~L., Struck, C. and Hancock, M.: 2010,
\journal{Astron. J.}, \vol{139}, 1212.

Stanimirovic, S., Staveley-Smith, L., Dickey, J.~M., Sault, R.~J.
and Snowden, S.~L.: 1999, \journal{Mon. Not. R. Astron. Soc.},
\vol{302}, 417.

Stanimirovi{\'c}, S. and Lazarian, A.: 2001, \journal{Astrophys.
J.}, \vol{551}, L53.

Stanimirovi{\'c}, S., Putman, M., Heiles, C., Peek, J. E. G.,
Goldsmith, P. F., Koo, B.-C., Krco, M., Lee, J.-J., Mock, J.,
Muller, E., Pandian, J. D., Parsons, A., Tang, Y., Werthimer, D.:
2006, \journal{Astrophys. J.}, \vol{653}, 1210.

Stanimirovi{\'c}, S., Hoffman, S., Heiles, C., Douglas, K.~A.,
Putman, M. and Peek, J.~E.~G.: 2008, \journal{Astrophys. J.},
\vol{680}, 276.

Staveley-Smith, L., Kim, S., Putman, M. and Stanimirovi{\'c}, S.:
1998, \journal{Rev. Mod. Astron.}, \vol{11}, 117.

Sternberg, A., McKee, C.~F. and Wolfire, M.~G.: 2002,
\journal{Astrophys. J. Suppl. Series}, \vol{143}, 419.

Taylor, C.~L., Walter, F. and Yun, M.~S.: 2001,
\journal{Astrophys. J.}, \vol{562}, L43.

Thilker, D.~A. et al.: 2009, \journal{Nature}, \vol{457}, 990.

T{\"u}llmann, R., Breitschwerdt, D., Rossa, J., Pietsch, W. and
Dettmar, R.-J.: 2006, \journal{Astron. Astrophys}, \vol{457}, 779.

Vieser, W. and Hensler, G.: 2007, \journal{Astron. Astrophys.},
\vol{472}, 141.

Walter, F. and Heithausen, A.: 1999, \journal{Astrophys. J.}, \vol{519}, L69.

Weilbacher, P.~M., Duc, P.-A., Fritze v.~Alvensleben, U., Martin,
P. and Fricke, K.~J.: 2000, \journal{Astron. Astrophys.},
\vol{358}, 819.

Werk, J.~K., Putman, M.~E., Meurer, G.~R., Oey, M.~S., Ryan-Weber,
E.~V., Kennicutt, R.~C., Jr. and Freeman, K.~C.: 2008,
\journal{Astrophys. J.}, \vol{678}, 888.

Wolfire, M.~G., McKee, C.~F., Hollenbach, D., Tielens, A.: 1995,
\journal{Astrophys. J.}, \vol{453}, 673.

}
\end{multicols}



\naslov{STRUKTURA NA MALIM SKALAMA MAGELANOVOG POTOKA KAO OSNOVA ZA GALAKTIQKU 
EVOLUCIJU}


\authors{S. Stanimirovi\'c, J. S. Gallagher III and L. Nigra}

\vskip3mm


\address{Astronomy Department, University of Wisconsin, Madison, 475 North
Charter Street, \break Madison, WI 53711, USA}

\Email{sstanimi}{astro.wisc.edu}


\vskip.7cm



\vskip1mm

\centerline{\rit Pregledni rad po pozivu}

\vskip.7cm

\begin{multicols}{2}
{


\rrm 

Magelanov potok (MP) je naj\-bli\-{\zz}i primer gasnog traga formiranog 
me\-{\dd}u\-dejst\-vom galaksija. Kako je znaqajna masa gasa u ovim vrstama 
okogalaktiqkih struktura pretpostavljena da predstavlja va{\zz}an izvor goriva 
za budu{\cc}e formiranje zvezda, mehanizam koji ovaj materijal mo{\zz}e vratiti 
nazad u galaksiju ostaje nejasan. Nedavna posmatranja neutralnog vodonika 
({\rm HI}) su pokazala da se\-ver\-ni deo MP, koji verovatno interaguje sa 
vrelim gasom iz haloa Mleqnog Puta blizu 1000 {\rm Myr}, zahvata ve{\cc}i 
prostor nego xto je prethodno bilo utvr{\dd}eno, mada tako{\dd}e sadr{\zz}i 
znaqajnu koliqinu struktura na maloj skali. Posle kratkog razmatranja 
kinematike na velikim skalama u MP uslovljenom nedavno otkrivenom proxirenosti 
MP, mi izuqavamo starost procesa u gasu MP kroz dejstvo raz\-li\-qi\-tih 
hidrodinamiqkih nestabilnosti i me{\dd}uzvezdanih turbulencija. Ovo nas vodi 
do razmatranja procesa u kojima materijal MP pre{\zz}ivljava kao hladna gasna 
faza, i jox uvek evidentno ne uspeva da formira zvezde. Pa\-ra\-le\-le 
izme{\dd}u MP i vangalaktiqkih plimskih formacija su kratko diskutovane sa 
nag\-la\-xa\-va\-njem koraka koji uspostavljaju xta MP otkriva o pravilima 
vezanim za lokalne procese u odre{\dd}ivanju evolucije sistema ovakve vrste. 
} 

\end{multicols}

\end{document}